\documentclass[aps,prl,twocolumn,superscriptaddress]{revtex4}

\usepackage{graphicx}
\usepackage[german,english]{babel}
\newcommand{\ang}{\ensuremath{\text{\AA}}}
\newcommand{\diff}{\ensuremath{\text{d}}}

\renewcommand{\Re}{\ensuremath{{\rm Re}\,}}
\renewcommand{\Im}{\ensuremath{{\rm Im}\,}}

\begin{document}

\title{Spectral functions of isolated Ce adatoms on paramagnetic surfaces}

\author{S. Gardonio}
\affiliation{Sincrotrone Trieste, Area Science Park, 34149 Trieste, Italy}
\author{T. O. Wehling}
\affiliation{I. Institut f{\"u}r Theoretische Physik, Universit{\"a}t Hamburg, Jungiusstra{\ss}e 9, D-20355 Hamburg, Germany} 
\author{L. Petaccia} 
\author{S. Lizzit}
\author{P. Vilmercati} 
\author{A. Goldoni} 
\affiliation{Sincrotrone Trieste, Area Science Park, 34149 Trieste, Italy}
\author{M. Karolak}
\author{A. I. Lichtenstein}
\affiliation{I. Institut f{\"u}r Theoretische Physik, Universit{\"a}t Hamburg, Jungiusstra{\ss}e 9, D-20355 Hamburg, Germany}
\author{C.Carbone}
\affiliation{Istituto di Struttura della Materia, Consiglio Nazionale delle Ricerche, Trieste, Italy}


\date{\today}

\begin{abstract}
We report photoemission experiments revealing the full valence electron spectral function of Ce adatoms on Ag(111), W(110) and Rh(111) surfaces. A transfer of Ce 4f spectral weight from the ionization peak towards the Fermi level is demonstrated upon changing the substrate from Ag(111) to Rh(111). In the intermediate case of Ce on W(110) the ionization peak is found to be split. This evolution of the spectra is explained by means of first-principles theory which clearly demonstrates that a reliable understanding of magnetic adatoms on metal surfaces requires simultaneous low and high energy spectroscopic information.
\end{abstract}
\maketitle
Ce atoms coupled to a metallic host present a characteristic case of quantum impurity problems: The competition of strong Coulomb interactions of the Ce 4f electrons and hybridization effects puts these states at the borderline between localized and delocalized. Hence, the physics of Ce compounds or Ce impurities in metallic hosts ranges local moment behavior in weakly hybridizing environments to non-magnetic Ce in the case of strong hybridization and rich Kondo physics in between \cite{HewsonBook,Allen_Kondo_Review}. To date, there are two experimental techniques allowing to probe the spectral properties of such systems: scanning tunneling spectroscopy (STS) and photoemission (PE) techniques (see \cite{Heinreich_JPhys09} and \cite{Allen_Kondo_Review} for recent reviews.)

Advances in nanotechnology like the ability of controlled atom manipulation, have put the electronic properties magnetic nanosystems into the focus of intense research: STS is well suited to study individual magnetic atoms on surfaces within a spectral range of some few $100$\, meV around the Fermi level and with very high spatial resolution. Starting in 1998, these techniques revealed how magnetic adatoms give rise to a sharp Abrikosov-Suhl resonance in the vicinity of the Fermi level --- a characteristic fingerprint of the Kondo effect \cite{Madhavan1998s,Berndt_98}. However, higher energy spectral properties ($4f$ or $3d$ ionization peaks and/or corresponding crystal-field-split states) are difficult to obtain with STS and one is restricted to the analysis of the Kondo resonance close to the Fermi energy. A test of the resulting Anderson model parameters with higher-lying levels is hardly possible. PE, on the contrary, can resolve spectra in the full range of the valence band but has so far been limited mainly to the study of Ce alloys (see \cite{Baer_Kondo_Review,Allen_Kondo_Review} for reviews). Measuring the full spectral properties of isolated magnetic adatoms on transition metal surfaces remains, therefore, an open experimental challenge.

In this letter, we extend the energy range in which the spectral properties of magnetic adatoms on metallic surfaces can be probed. We report on photoemission experiments revealing the full valence band electronic structure of isolated Ce adatoms on the surfaces of Ag(111), W(110) and Rh(111) and fully describe the valence band electronic structure of isolated Ce adatoms on different these hosts. For the first time, our studies reveal the $4f$  ionization peaks and the Kondo excitations for isolated Ce adatoms on a metallic surface in a single experiment. By means of first-principles calculations we explain the photoemission results and yield the link between the atomistic environment of the Ce adatoms and their spectroscopic properties.

We prepared the Ag(111), W(110) and Rh(111) substrates by the standard procedures and monitored their crystalline quality by low energy electron diffraction (LEED). Isolated adatoms were obtained by depositing Ce atoms at a substrate temperature of 20K in a coverage regime ranging from one thousandth to few tenth of a monolayer (statistical growth regime).
A major technical difficulty in photoemission experiments on isolated magnetic adatoms on metal surfaces (requiring adatom coverage in the range of 0.01-0.001 ML) is that the spectroscopic signature of the adatoms can be obtained only for those combinations of impurities and substrates for which the signal of the impurity is comparable to the host. We demonstrate that this condition is fulfilled for photoemission measurements of $4f$ valence states of rare earth isolated atoms on transition and noble metal surfaces measured across the $4d\to 4f$ threshold (resonant photoemission). Here, the required sensitivity to the adatoms is achieved because of the enhanced photoemission cross section of the $4f$ states as compared to the $3d$, $4d$ or $5d$ photoemission cross sections of the transition and noble metals substrates.  The photoemission experiments were performed at SuperESCA beamline at ELETTRA Synchrotron Radiation facility. We measured photoemission energy distribution curves of the Ce 4f valence state at 122 eV photon energy, which corresponds to Ce $4d\to 4f$ resonance, with an overall energy resolution of $40$\,meV. 

\begin{figure}
\centering
\includegraphics[width=.95\linewidth]{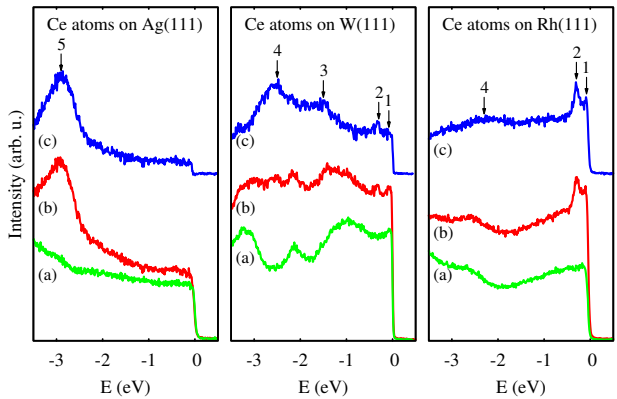}
\caption{\label{fig:Ce_exp_spectra}(Color online) Valence photoelectron energy distribution curves for the clean Ag(111), W(110), and Rh(111) surfaces (a) and in presence of Ce adatoms (~0.001 ML coverage) (b). The curves (c) are the differences between the corresponding spectra (b) and (a).}
\end{figure}

Fig. \ref{fig:Ce_exp_spectra} shows the resonant photoelectron spectra measured at the Ce $4d\, \rightarrow 4f$ absorption threshold for Ce isolated atoms on Ag(111), W(110) and Rh(111) surfaces. 
In order to better distinguish between the photoemission signal derived from the Ce adatoms and the substrate contribution, the photoemission spectra before (a) and after (b) deposition of the Ce atoms are compared and the difference spectra (c) are taken for each host. 

The spectra of isolated Ce atoms are found to be markedly different on the three substrates and a clear trend when going from Ag to Rh is visible: On Ag(111) only a peak around -2.9 eV is present, while on W(110) peaks at energies of -1.5 and -2.3 eV as well as two smaller peaks close to the Fermi level, at 0 and -0.25 eV, contribute spectral weight. On Rh(111) the major contribution is derived from structures close to the Fermi level, the peaks labelled 1 and 2, and a small broad peak around -2.3 eV is still present. The $4f$ electrons of Ce seem to undergo a transition from localized to delocalized when going from Ce on Ag (111) to Ce on Rh (111) via the intermediate case of Ce on W (110) where the ionization peak appears to be split up into two peaks.

To understand the origin of these entirely different distributions of Ce $4f$ spectral weight and to explain why the ionization peak is split up on W (110), we proceeded in two steps. First, density functional theory (DFT) calculations are performed to obtain relaxed geometries and the hybridization functions for single Ce-atoms on Ag(111), W(110), and Rh(111). These hybridization functions are afterwards used to calculate the valence electron photoemission spectra following Gunnarsson and Sch{\"o}nhammer \cite{Gunnarsson_Schoenhammer83}. The DFT calculations have been carried out using the Vienna Ab-Initio Simulation Package (VASP) \cite{Kresse:PP_VASP} with the projector augmented waves basis sets (PAW) \cite{Kresse:PAW_VASP,Bloechl:PAW1994}. In these calculations, single Ce atoms on Ag(111), W(110) and Rh(111) have been modeled using $2\times 2$ surface supercells with slab thicknesses of 7 layers. The crystal structures have been relaxed until the forces acting on each atom were less than 0.02eV/$\ang$. In all cases, Ce adsorbs to high symmetry positions, which are close to the continued bulk lattice on Ag(111) and W(110). On the Rh (111) surface, Ce adsorption to an hcp site is by $\sim 250$\,meV more favorable than the fcc site.

The Ce 4f electrons are expected to be subject of strong correlations which require a description in terms of an LDA++ approach \cite{Lichtenstein_LDA+DMFT_98,DFT++}: For the description of the Ce adatoms we choose Anderson impurity models with the hybridization functions, $\Delta_{mm'}(\omega)$, being obtained from DFT. Here, $m,m'=-3,...,3$ label the $z$-component of the angular momentum of the Ce $4f$ orbitals.
The PAW basis sets provide intrinsically projections onto localized atomic orbitals, which we use to extract the hybridization functions (for details see \cite{PAW_DMFT,Co_graphene_FanoPRB,DFT++}).

\begin{figure}
\includegraphics[width=.98\linewidth]{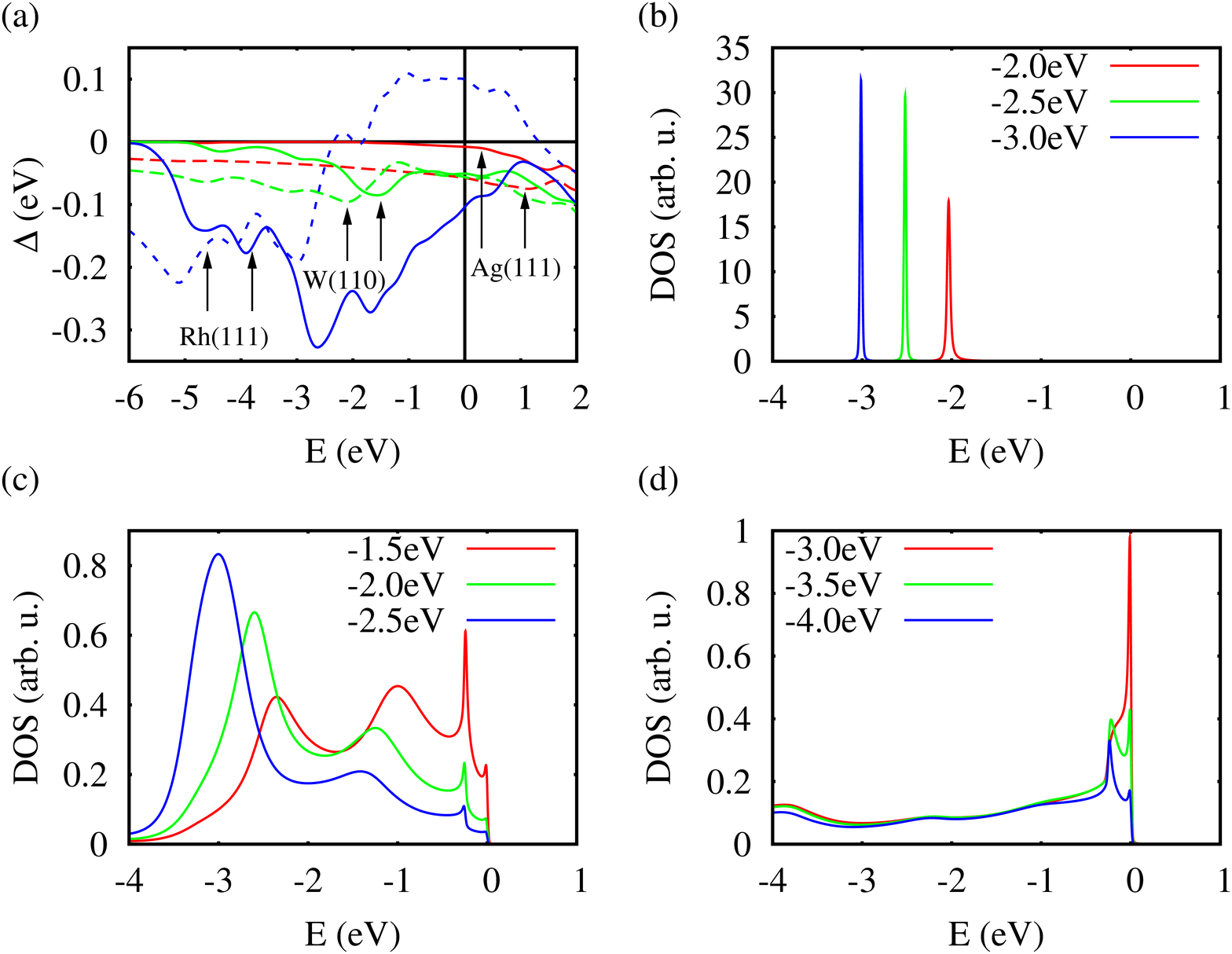}
\caption{\label{fig:Ce_hyb_spec}(Color online) (a) Calculated Ce 4 f-electron hybridization functions for isolated adatoms on Ag(111) (red), W(110) (green), and Rh(111) (blue). The real parts are shown as dashed and the imaginary parts as solid lines. (b-d) Calculated $4f$-spectra for single Ce atoms on Ag(111) (b), W(110) (c) and Rh(111) (d) surfaces. The dependence of the spectra on the on-site energies $\epsilon_f$ is illustrated by showing spectra obtained with different $\epsilon_f$. All curves are labelled by the corresponding  $\epsilon_f$.}
\end{figure}

These hybridization functions act like energy dependent complex valued potentials on the Ce f-orbitals and fully characterize their interaction with the substrate as regards local observables like the Ce f-spectral functions probed in the photoemission experiments. Following Ref \cite{Gunnarsson_Schoenhammer83}, we assume $\Delta_{mm'}(\omega)=\Delta(\omega)\delta_{mm'}$, which we obtain from the full LDA hybridization, $\Delta^{\rm LDA}_{mm'}(\omega)$ by averaging over all Ce f-orbitals, $\Delta(\omega)=\frac{1}{7}\sum_m\Delta^{\rm LDA}_{mm}(\omega)$. The resulting $\Delta(\omega)$ for Ce on the three different substrates are shown in Fig. \ref{fig:Ce_hyb_spec} (a).

The imaginary part of the Ce 4f-hybridization, $\Im\,\Delta(\omega)$, on the Ag (111) is at all energies below the Fermi level more than an order of magnitude smaller than the corresponding 4f-hybridizations of Ce on W(110) or Rh (111). Obviously, Ce on Rh (111) exhibits by far the strongest hybridization, which gives the first hint, why the ionization peak of Ce on Rh (111) is more strongly smeared out and why the quasi-particle peak at the Fermi level might be strongest for Ce on Rh (111). In the intermediate case of Ce on W(110), there is a peak in $\Im\,\Delta(\omega)$ extending from $\omega=-1$\,eV to $-2$\,eV. This resonance is caused by bonding to $t_{2g}$ $d$-orbitals of neighboring W-atoms.

The real parts of the Ce $4f$-hybridization functions are related to $\Im\Delta(\omega)$ by the Hilbert transform $\Re\Delta(\omega)=-\frac{1}{\pi}\mathcal{P}\int\diff\omega' \frac{\Im\Delta(\omega')}{\omega-\omega'}.$ Hence, peaks in the imaginary part cause steps in the real part. The step in $\Re\,\Delta(\omega)$ in the energy range of $-2.1$\,eV to $-1.3$\,eV for Ce on W(110) is responsible for the peak splitting observed in photoemission (see Fig. \ref{fig:Ce_exp_spectra}).

Using Eqs. (A1-A7) from Ref. \onlinecite{Gunnarsson_Schoenhammer83}, we obtain the ground state properties of the Ce 4f electrons including the energy gain $\delta$ upon hybridization and the average occupancy $n_f$ as well as the valence electron photoemission spectra from the hybridization functions. Here, only the on-site f-energy remains as a fitting parameter, while we use the established value of $250$\,meV for the spin-orbit splitting between the $f_{5/2}$ and $f_{7/2}$ states.

Fig. \ref{fig:Ce_hyb_spec} (b-d) shows the calculated Ce $4f$ spectra for all substrates and different on-site energies $\epsilon_f$. The spectra of Ce on Ag(111) exhibit one sharp peak close to the bare $f$-electron energy, which corresponds to the ionization process $f^1\rightarrow f^0$. Independent of $\epsilon_f$ the spectral weight of this main peak dominates the spectrum and is by orders of magnitude larger than the weight of a possible Kondo peak: within the entire range of $\epsilon_f$ considered in Fig. \ref{fig:Ce_hyb_spec} (b-d) we find $1-n_f<10^{-4}$. This represents a measure for the weight $Z$ of the Kondo peak, as $Z=1-n_f$ for vanishing spin-orbit splitting $\Delta \epsilon_f=0$ \cite{Gunnarsson_Schoenhammer83,HewsonBook}.

For the case of Ce on Rh(111), the Kondo or the spin-orbit peak ($f_{5/2}\rightarrow f_{7/2}$ around the energy $-\Delta\epsilon_f$) dominates the spectrum for all $\epsilon_f$ shown in Fig. \ref{fig:Ce_hyb_spec} (b-d) and the ionization peak is always broadened to a wide continuum. Here, the main effect of changing $\epsilon_f$ is changing the relative weight of the spin-orbit peak as compared to the Kondo peak.

The spectra of Ce on W(110) lie in between these two extreme cases with the ionization peak being remarkably split-up into two separate peaks at $E>-1.5$\,eV and $E< -2.3$\,eV in all cases. In this intermediate regime, the on-site-energy strongly affects the ratio of spectral weight in the ionization peak to the amount of weight in the Kondo and the spin-orbit peaks.

\begin{figure}
\centering
\includegraphics[width=.9\linewidth]{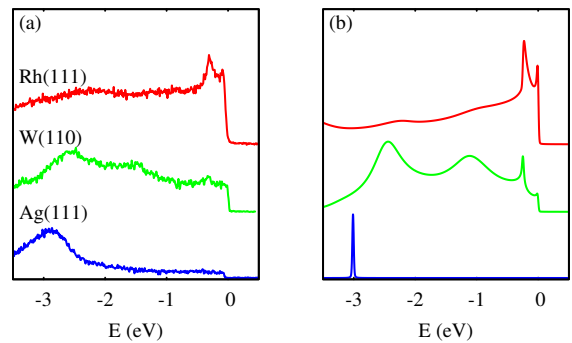}
\caption{\label{fig:Ce_exp_theo_comp}(Color online) Comparison of experimental (a) and theoretical spectra (b). The theoretical spectra present best fits to the experiment obtained with $\epsilon_f=-3$eV, $-1.7$eV, and $-3.7$eV for Ce on Ag (111), W (110) and Rh (111), repsectively.}
\end{figure}

The comparison of experimental and theoretical spectra (Fig. \ref{fig:Ce_exp_theo_comp}) shows that theory and experiment are in very good agreement.  This suggests that the physical processes determining the photoemission spectra of isolated Ce adatoms on the different surfaces are well described within our model. All trends including the peak splitting in the case of Ce on W(110) are understandable from the structure of the hybridization functions shown in Fig. \ref{fig:Ce_hyb_spec} (a). On the ionization peak, the hybridization acts like a self-energy term on a resonant level with the real part shifting the level and the imaginary part broadening it. The Ag(111) and Rh(111) surfaces are forming the two extreme cases of weak and strong hybridization, respectively, and lead to correspondingly sharp / broad ionization peaks. The resonance near $\omega=-1.6$\,eV in the hybridization of Ce on W(110) splits the ionization peak: the step in the real part causes spectral weight being shifted to lower energies below the resonance and to higher energies above it.

For Ce on Rh (111), there is an even larger step in $\Re\,\Delta(\omega)$ extending from $\omega=-3$\,eV to $-2$\,eV. This large step results in a significant amount of weight being pushed to that high energies, that it contributes to the peaks close to the Fermi level but does not form a second separate ionization peak as in the case of Ce on W(110). This effect occurring in the case of strong hybridization, has been similarly observed in the bulk compound CeNi$_2$ \cite{Allen_86,HewsonBook}.

As in the case of bulk alloys \cite{Allen_86}, the hybridization of the Ce 4f states with the conduction electron states can induce strong structures in the Ce 4f spectral function for isolated atoms on surfaces.  For the case of Ce on Ag (111), photoemission and theory found concordantly no significant Kondo like contribution to the spectra. This is in contrast to the Kondo effect of Ce on Ag (111) reported in early STM experiments \cite{Berndt_98}. The Kondo temperature $T_K=5$\,K$=0.4$\,meV$/k_B$ reported there is out of the range $T_K\sim\delta \ll 10^{-2}$meV for all $\epsilon_f<1$\,eV found here by combining photoemission and first-principles calculations. Hence, our results support the conjecture (c.f. Ref. \cite{Heinreich_JPhys09}) that the STM experiments from Ref. \cite{Berndt_98} have been measuring Ce clusters on Ag (111).

In conclusion, we demonstrated the ability of photoemission spectroscopy to measure the full valence electron spectral function of \textit{isolated} Ce adatoms on different metallic surfaces. By comparing the experimental results with first-principles calculations of the photoemission spectra for the atoms deposited on Ag (111), W(110) and Rh (111) surfaces, we find delocalization of the Ce $4f$ electrons when going from Ag to Rh substrates and observe the shift spectral weight from the ionization peaks to quasiparticle resonances at the Fermi level. The intermediate case of Ce on W(110) shows that the energy dependence of the impurity hybridization with the surface determines the shape of excitation spectra. Widening the energy window for probing the electronic structure of magnetic nanosystems appears to be a powerful way to disentangle hybridization mechanisms and understand the physics of magnetic nanosystems down to the atomic level.

Support from SFB 668 (Germany) as well as computer time at HLRN are acknowledged.
\bibliography{Ce_metal}

\end{document}